\documentclass[showpacs,twocolumn,preprintnumbers,prl,amssymb,floatfix]{revtex4}
\usepackage{graphicx}
\usepackage{dcolumn}
\usepackage{bm}
\usepackage{amsmath}
\usepackage{amsfonts}
\usepackage{amssymb}
\usepackage{ulem}
\usepackage{color}
\usepackage[T1]{fontenc}

\begin{document}
\title{Unveiling the Landau Levels Structure of Graphene Nanoribbons}
\author{Rebeca Ribeiro$^{1}$}
\author{Jean-Marie Poumirol$^{1}$}
\author{Alessandro Cresti$^{2}$}
\author{Walter Escoffier$^{1}$}
\author{Michel Goiran$^{1}$}
\author{Jean-Marc Broto$^{1}$}
\author{Stephan Roche$^{3,4}$}
\author{Bertrand Raquet$^{1}$}
\affiliation{$^{1}$Laboratoire National des Champs Magn\'{e}tiques
Intenses, INSA UPS CNRS, UPR 3228, Universit\'{e} de Toulouse, 143
av. de Rangueil, 31400 Toulouse, France}
\affiliation{$^{2}$ IMEP-LAHC (UMR CNRS/INPG/UJF 5130), Grenoble INP Minatec, 3 Parvis Louis N\'eel, BP 257, F-38016 Grenoble, France}
\affiliation{$^{3}$CIN2 (ICN-CSIC) and Universitat Autonoma de Barcelona, Catalan Institute of Nanotechnology, Campus de la UAB, 08193 Bellaterra (Barcelona), Spain}
\affiliation{$^{4}$ ICREA, Institucio Catalana de Recerca i Estudis Avan\c cats, 08010 Barcelona, Spain}
\date{\today}

\begin{abstract}
Magnetotransport measurements are performed in ultraclean
(lithographically patterned) graphene nanoribbons down to 70 nm.
At high magnetic fields, a fragmentation of the electronic
spectrum into a Landau levels pattern with unusual features is
unveiled. The singular Landau spectrum reveals large
magneto-oscillations of the Fermi energy and valley degeneracy
lifting. Quantum simulations suggest some disorder threshold at
the origin of mixing between opposite chiral magnetic edge states
and disappearance of quantum Hall effect.
\end{abstract}

\maketitle

{\it Introduction}.- To benefit from the unusual transport
properties of graphene (massless Dirac fermions
physics)~\cite{Geim,Castro}, and engineer novel building blocks
for future carbon-based nanoelectronics~\cite{NN}, the fabrication
of clean materials has become a central issue. Of great concern is
the design of graphene nanoribbons (GNRs) with lateral sizes in
the order of a few to tens of nanometers, which allow some gap
engineering~\cite{Han07}. The transverse confinement leads to 1D
electronic sub-bands profiles whose details depend on the width
and the edge geometry of the ribbons~\cite{Cresti}.

In presence of a strong perpendicular magnetic field, anomalous
quantum Hall effect develops in two-dimensional graphene with
unique properties widely discussed in the literature~\cite{QHED}.
The electronic spectrum is theoretically predicted to evolve into
magneto-electronic sub-bands resulting from a competition between
magnetic and electronic confinement~\cite{Castro,Cresti}. This is
partly unveiled by anomalous Shubnikov-de Haas oscillations when
the Landau diameter becomes larger than the ribbon's width
~\cite{PeresBerger}. However, it is puzzling to note the lack of
experimental evidence of Hall quantization in nanoribbons narrower
than $200$ nm \cite{Molitor,Oostinga,Poumirol,NNMR2}. Recent
magnetotransport experiments in chemically derived narrow GNRs
reported some signatures of chiral magnetic edge states revealed
by a large positive magnetoconductance~\cite{Poumirol}. Positive
magnetoconductance is also observed in lithographic
nanoribbons~\cite{Oostinga,NNMR2}, but in all these experiments,
the conductance is far from being quantized, and transport remains
strongly diffusive, thus jeopardizing a convincing observation of
the underlying Landau levels. Several sources of disorder are
suspected to crosslink chiral edge currents for the narrowest
ribbons, thus preventing quantum Hall effect from
developing~\cite{GNRdis}. Additionally, for clean GNRs, recent
theoretical calculations suggest a strong impact of
electron-electron interactions on the band structure leading to
the suppression of the conductance quantization~\cite{GNRee}. A
detailed characterization of Landau levels in GNRs thus remains to
be accomplished.

In this Letter, we report two-terminal quantum Hall resistance
measurements on GNR devices. The oscillatory behavior of the
magnetoresistance (MR) evidences a clear signature of
magneto-electronic subbands. The onset of a singular graphene-like
Landau spectrum reveals a large pinning of the Fermi energy along
with a valley degeneracy lifting. To rationalize those features,
we simulate the spatial extension of the corresponding magnetic
edge states and their distribution in presence of disorder.

\begin{figure} [!ht]
\begin{center}
\includegraphics[bb=16 16 162 147, width=0.7\linewidth]{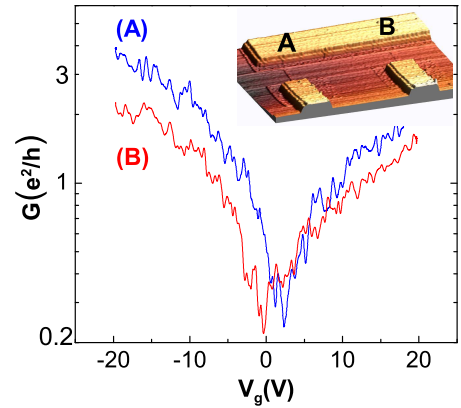}
\caption{(color online) Experimental $G(V_{\rm g})$ curves measured at
4.2K on two GNR devices of width 100 and 70nm (sample A and B).
Inset : the AFM image of the devices.} \label{fig0}
\end{center}
\end{figure}

Graphene devices are made by mechanical exfoliation of graphite
onto $n^{++}$Si/SiO$_2$(300nm) substrates. Electrodes are defined
by e-beam lithography followed by thermal evaporation of
Ti(1nm)/Pd(10nm)/Au(40nm). Next, the graphene flakes are patterned
into ribbons by oxygen plasma etching, using PMMA as an etching
mask. A set of connected GNRs are prepared with width ($W$)
ranging from 60 to 100nm and length ($L$) from 350 to 800nm. In
the following, we present extended results on two devices (inset
Fig.1), sample A ($L=350$nm, $W=100$nm) and B ($L=750$nm,
$W=70$nm), having the hallmarks of the overall samples.

\begin{figure} [!ht]
\begin{center}
\includegraphics[bb=16 16 281 223, width=0.85\linewidth]{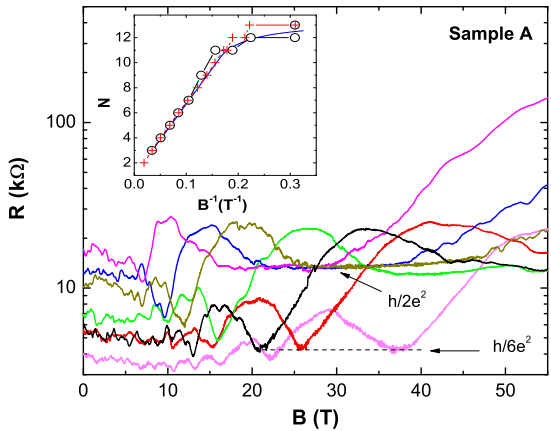}
\caption{(color online) Two-probe perpendicular magnetoresistance
measured at 4.2K on sample A, for selected $V_{\rm g}$ from -10V
(cyan) to -40V (light magenta), by step of 5V. The inset shows
anomalous $1/B$ Shubnikov-de Haas oscillations with circle marks
indicating the Landau index $N$ as a function of the $1/B_i$
locations deduced from the MR curve at -40V. The red crosses are
the $N$,($1/B_i$) simulated data from the band structure
(Fig.~\ref{fig3}a, inset). The blue curve is the calculated
$N$($1/B_i$) from~\cite{Sup}.} \label{fig1}
\end{center}
\end{figure}

After standard thermal annealing treatment in vacuum to desorb
contaminants, the measured $G(V_{\rm g})$ at 4.2K
(Fig.~\ref{fig0}) exhibits a minimum at the charge neutrality
point corresponding to a back-gate potential $V_{\rm g}=V_{\rm
CNP}$, close to zero, $2.5$V and $-0.5$V for samples A and B,
respectively, thus pinpointing a negligible residual doping. From
a numerical calculation of the electrostatic coupling between the
ribbons and the back-gate (available at
http://www.fastfieldsolvers.comm), the carrier density is
estimated as $n({\rm m}^{-2}) \approx 1.5\times10^{15} \times
(V_{\rm g}({\rm V})-V_{\rm CNP})$ while the Fermi energy scales as
$E_F$(meV) $\approx 40 \times \sqrt{(V_{\rm g}({\rm V})-V_{\rm
CNP})}$. Assuming a negligible contribution of quantum
interferences, we estimate the electronic mean free path,
$l_m(V_{\rm g})$, from the $G(V_{\rm g})$ curves ~\cite{Sup}. At
$V_{\rm g}=-20$V, we infer $l_m\approx 50-120$nm (80-120nm) for
sample A(B). In both cases, $l_m \sim W$ and $L/l_m \approx 2-7$,
this indicates that the transport regime is close to a
quasi-ballistic regime. The  good quality of the samples is also
confirmed by the estimated large field effect mobility at 4.2K $
\mu_{{\rm A}({\rm B})} \approx 1200(3500)$cm$^2$/(Vs), and by the
Fabry-Perot conductance modulations observed at 2K~\cite{Sup}.

\begin{figure}[h!]
\begin{center}
\includegraphics[bb=16 16 181 168,width=0.8\linewidth]{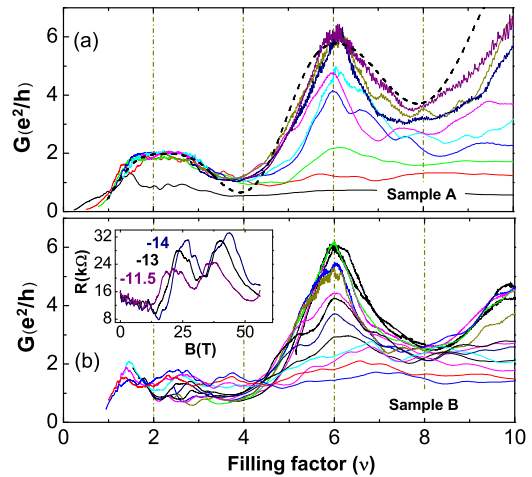}
\caption{(color online) (a) Experimental magnetoconductance of
sample A versus the filling factor deduced from the MR curves at
selected $V_{\rm g}$, from -40V (top) to 0V (bottom), by step of
5V. The dashed line is the simulated $G(\nu)$.(b)
Magnetoconductance of sample B versus the filling factor at
different $V_{\rm g}$ from -50V (top) to 0V (bottom), by step of
5V. Inset: the singular MR curves exhibiting a double resistance
peak when $E_F$ crosses the $N=$2 Landau level.} \label{fig2}
\end{center}
\end{figure}

Under a perpendicular magnetic field, the two-terminal resistance
shows a non-trivial sample-shape dependent profile. It reveals
fingerprints of both the quantized Hall resistance for
incompressible charge carriers densities (i.e. for filling factors
$\nu(n_eh/eB)=4(n+1/2)$) and the longitudinal resistance for
intermediate filling factors \cite{WilliamsAbanin}. Accordingly,
for elongated ribbons, the resistance peaks pinpoint the
depopulation onset of the highest occupied Landau subband of index
$N$. Figure~\ref{fig1} shows the two-probe MR up to 55T at various
$V_{\rm g}$ for sample A. An oscillatory behavior of the
resistance along with quantized minima ($h/6e^2$) and resistance
plateaus ($h/2e^2$) are clearly visible. The MR plot as a function
of the inverse magnetic field shows the expected $1/B$
Shubnikov-de Haas oscillations of the resistance peaks for a 2D
electron gas~\cite{Sup}.

\begin{figure}[h!]
\includegraphics[bb=17 18 170 232,width=0.9\linewidth]{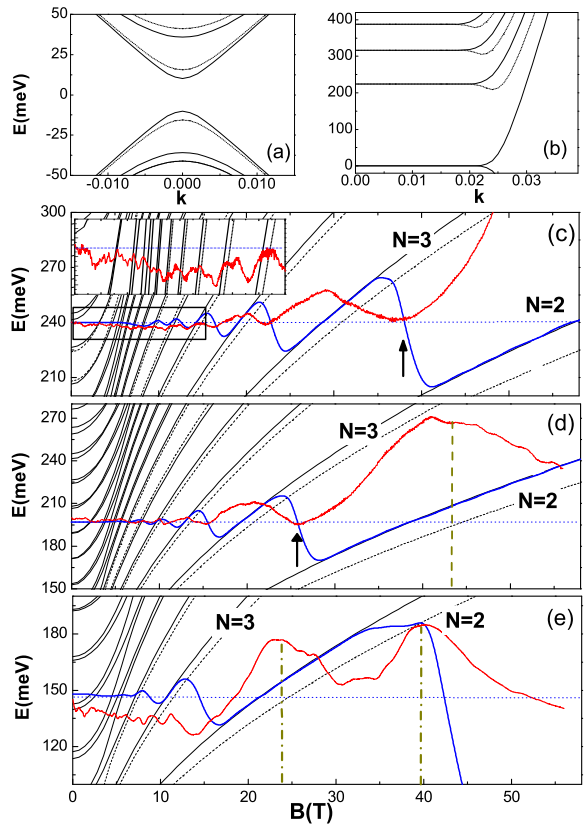}
\caption{(color online) (a) and (b) Band structure of a 70nm-wide
aGNR around the CNP at 0 and 50T, respectively. (c-e) Simulated
magneto-electronic subbands versus magnetic field for the
100nm-wide aGNR (c) and 70nm-wide aGNR (d,e) directly compared to
quantum oscillations at selected $E_F$. Black lines hold for the
edge subbands at zero-$k$. The dashed black lines correspond to
the valley degeneracy lifting at nonzero $k$. In red are plotted
the MR curves in arbitrary unit and in blue, the related Fermi
energy. Dark yellow vertical lines indicate the resistance peaks
shifted to higher magnetic field due to the Fermi energy pinning
at low $N$.} \label{fig3}
\end{figure}

However, a strong departure from the $1/B$ periodicity is observed
for larger $N$ (inset, Fig.~\ref{fig1} circle marks). This is a
convincing signature of the electronic confinement that starts to
overcome the magnetic one when the cyclotron radius becomes larger
than $W/2$. The linearity and its deviation above $N \approx 9$
are well reproduced by the calculation of the number of occupied
sub-bands $N=N(1/B_i)$ in the frame of semiclassical
Bohr-Sommerfeld quantization rule with an hard-wall confinement
(blue line)~\cite{Sup}.

Figure~\ref{fig2} shows the conductance profiles of the two GNRs
as a function of the filling factor. The curves are deduced from
the MR experiments, i.e. a constant charge density along with a
pulse field sweep, meaning an increase of the Landau energy
broadening versus $\nu$. For sample A (Fig.2a), a clear $2G_0$
Hall conductance plateau is observed at $\nu = 2$, providing
direct evidence of a single layer graphene. For larger
incompressible charge densities, at $\nu = 6, 10,...$, only maxima
of conductance develop instead of the expected $(6,10,...)G_0$
plateaus. This overall behavior is consistently described in the
context of the conformal invariance of the conductance, where the
distortion of the plateau is driven by the device aspect ratio $
\xi = L/W$. Following the theory developed in
~\cite{WilliamsAbanin}, we reasonably well simulate our data for $
\xi_{fit} = 4.1$ (to be compared to $ \xi_{exp} = 3.5$) and a
broadening of $\triangle \nu \approx 0.84$ and $1.05$ around $\nu
= 2$ and $6$, respectively (Fig.\ref{fig2}a, dashed line). The
larger broadening at higher filling factor is therefore
responsible for the shrinkage of the $6G_0$ quantized state.

Despite such agreement, several intriguing experimental features
demand further considerations : \textit{(i)} The resistance peaks
preceding the $h/2e^2$ plateau on sample A (i.e. corresponding to
the crossing of the $N=3$ Landau level) appear drastically
enlarged for the MR curves at larger $V_{\rm g}$ (Fig.\ref{fig1}).
\textit{(ii)} Surprisingly, the Hall conductance of sample B,
which is 30\% narrower, does not present a well-defined $2G_0$
plateau even though the $6G_0$ quantization at $\nu = 6$ is
preserved (Fig.\ref{fig2}b). The suppression of the $2G_0$
conductance plateau goes along with a well-marked splitting and
broadening of the resistance peak that develops before the
expected plateau (Fig.\ref{fig2}b-inset). \textit{(iii)} Both
samples exhibit a gradual suppression of the $6G_0$ conductance at
$\nu$=6 as $V_g$ decreases (Fig.\ref{fig2}a and b).

For a deeper understanding of the MR curves and the related Landau
levels pattern, we now consider the magnetic field dependent band
structure of two armchair ribbons (aGNRs) of width $W\approx100$
and $W\approx70$nm. Due to the rather large width of the ribbons,
the following discussion does not depend on the exact number of
dimer lines that compose the ribbons. Figures~\ref{fig3}(a,b) show
the band structure of the narrowest aGNR around the CNP at $B$=0
and 50T, respectively. Note that the valley degeneracy of the
two-dimensional graphene is lifted as a consequence of the ribbon
boundary conditions~\cite{Cresti}. At $B$=50T, the bulk Landau
levels are degenerated while the edge states (at large wavenumber
$k$), split into couples of nondegenerate bands, among which one
has an increasing energy while the other first decreases and then
rises when approaching the edges of the Brillouin zone. The black
(solid and dashed) lines of Fig.~\ref{fig3}(c-e) indicate the
minimum energy of each band as a function of the magnetic field.
Above 10T, they start to scale as $\sqrt{B}$, thus converging to
the Landau levels scaling of the 2-D graphene. Interestingly, the
degeneracy breaking of edge states is enhanced at high fields. To
relate the transport oscillations with the underlying band
structure, we compare the intersection of the Fermi energy and the
magneto-subbands spectrum with the locations of the maxima of
resistance (red curve). Assuming a constant $E_F$ (blue dashed
line), a good agreement between the locations of the resistance
peaks and the subband depopulation for high quantum numbers is
observed at low fields (inset of Fig.~\ref{fig3}a and
corresponding red crosses in the inset of Fig.~\ref{fig1}).
However, a noticeable mismatch gradually increases below $N$=3 and
reaches several Tesla for $N=2$ in Fig.~\ref{fig3}(d,e), marked by
vertical dashed lines~\cite{Sup}.

To shed light on this issue, we observe that, as in conventional
2D gas, the Fermi energy is not constant but varies to accommodate
the carriers into the available magnetic field dependent subbands
~\cite{Sup}. This determines large oscillations of the Fermi
energy at low $N$ (see the blue lines in Fig.~\ref{fig3}), with an
evident pinning at the Landau levels at high magnetic fields. The
Fermi energy oscillations provide a clear explanation of the shape
of the MR curves in the high field regime. Indeed, the inflexion
point of the $E_F(B)$ curves in between two successive Landau
levels well matches with the minima of resistance
(Fig.~\ref{fig3}(c,d), marked by arrows). Besides, the widening of
the resistance peaks at larger $V_{\rm g}$ and their shift to high
fields are a direct consequence of a stronger Fermi energy pinning
onto the lowest index Landau levels~\cite{Sup}. In this framework,
the absence for sample B of a well-defined $2G_0$ plateau
accompanying the splitting and the widening of the resistance peak
also finds a natural explanation (Fig.~\ref{fig3}e and
~\cite{Sup}): The second maximum of resistance coincides with the
changeover of $E_F$ from the $N=2$ Landau level (solid black line)
to the second energy minimum at nonzero $k$ (dashed black line).
Note that such a signature of the valley degeneracy lifting on the
MR curves requires a large enough energy splitting between the two
sub-levels along with a pinning of the Fermi energy on the two
states. These conditions are borne out on sample B, which has a
higher mobility and allows for a sharper valley degeneracy lifting
with an higher density of states at the subband edges.

\begin{figure}[h!]
\begin{center}
\includegraphics[bb=14 14 1694 1203,width=0.8\linewidth]{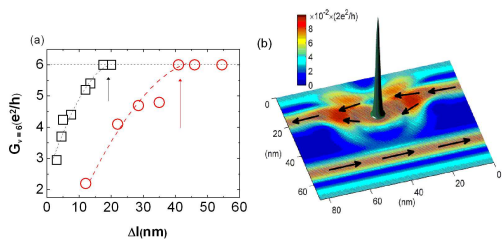}
\caption{(color online) (a) The measured suppression of the
quantized conductance at $\nu=6$ for sample A (red) and B (black)
versus the simulated current-free bulk width $\Delta l(nm)$. (b)
Spatial distribution of the background spectral currents through a
70nm-wide aGNR at $\nu = 6$ ($E_F=119$meV and $B=12$T), in
presence of a Gaussian potential with maximum strength $V=1$ eV,
mimicking a single charged impurity.} \label{fig4}
\end{center}
\end{figure}

We finally comment on the gradual suppression of the $6G_0$
conductance at $\nu = 6$ when decreasing the back-gate voltage
(Fig.~\ref{fig2}). The deterioration of the conductance
quantization at an incompressible filling factor can be related to
the presence of disorder that crosslinks the chiral currents
flowing at opposite edges. In ~\cite{Sup}, we simulate the spatial
profile, along the transverse section of a pristine 70nm-wide
aGNR, of the two counter propagating edge channels at $\nu = 6$ as
a function of the magnetic field. As the magnetic field increases,
the Landau states extend more and more in the bulk and the edge
states are progressively pushed toward the edges. We extract the
distance between the edge channels ($\Delta l$) for a given
magnetic field. Figure~\ref{fig4}a shows the measured conductance
at $\nu = 6$ as a function of $\Delta l$. The conductance starts
to degrade when the corresponding edge channels in the pristine
ribbon are still well separated by 40 and 15nm for the sample A
and B, respectively. This is consistent with the higher degree of
disorder of sample A, which is able to crosslinks the chiral
currents more effectively. This behavior is very sample-dependent,
since a specific disorder configuration might make the chiral
currents come significantly closer, thus facilitating
backscattering. Additionally, the lower $V_{\rm g}$ the more
pronounced the phenomenon is, owing to a reduced screening of the
impurity potential. As a matter of illustration, Fig.~\ref{fig4}b
shows the leakage of edge currents (at $N=6$) provoked by a single
charged impurity for a 70nm-wide graphene ribbon.

{\it Conclusion}.- Unusual features of Landau levels have been
explored in clean graphene nanoribbons using high magnetic field
transport measurements. The observed intriguing transport
oscillations have been related to Fermi level pinning and valley
degeneracy lifting. By means of quantum simulations, the spatial
extension and robustness to disorder of chiral edge currents have
been discussed.

{\it Acknowledgements}. -Sample preparations were achieved at
LAAS. Part of this work is supported by EuroMagNET, contract n$^o$
228043. S.R. acknowledges  the NANOSIM-GRAPHENE Project ${\rm
N.}^{\circ}$ ANR-09-NANO-016-01. A.C. acknowledges the support of
Fondation Nanoscience via the RTRA Dispograph project.


\begin{thebibliography}{50}
\bibitem{Geim}
A.K. Geim and K.S. Novoselov, Nature Materials {\bf 6}, 183 (2007).

\bibitem{Castro}
A.H. Castro Neto {\it et al.}, Rev. Mod. Phys. {\bf 81}, 109
(2009). 

\bibitem{NN}
F. Schwierz, Nature Nanotechnology {\bf 5 }, 487-496 (2010).

\bibitem{Han07}
M.Y. Han {\it et al.}, Phys. Rev. Lett. \textbf{98}, 206805 (2007).


\bibitem{Cresti}
K. Wakabayashi {\it et al.}, Phys. Rev. B. \textbf{59}, 8271
(1999); K. Wakabayashi, Phys. Rev. B \textbf{64}, 125428 (2001);
A. Cresti {\it et al.}, Nano Res. \textbf{1}, 361 (2008).

\bibitem{QHED}
K.S. Novoselov {\it et al.}, Nature {\bf  438}, 197 (2005); Y.B Zhang {\it et al.}, Nature {\bf  438}, 201 (2005); K. S. Novoselov{\it et al.},  Science  {\bf 315}, 1379 (2007).

%

\bibitem{PeresBerger}
N.M.R. Peres {\it et al.},  Phys. Rev. B \textbf{73}, 241403(R)
(2006); C. Berger {\it et al.}, Science \textbf{312}, 1191 (2006).

\bibitem{Molitor}
F. Molitor {\it et al.}, Phys. Rev. B \textbf{79}, 075426 (2009).

\bibitem{Oostinga}
J.B. Oostinga {\it et al.}, Phys. Rev. B \textbf{81}, 193408
(2010).

\bibitem{Poumirol}
J. Poumirol {\it et al.}, Phys. Rev. B \textbf{82}, 041413(R)
(2010)


\bibitem{NNMR2}
J. Bai {\it et al.}, Nature Nanotechnology {\bf 5}, 655-659 (2010).

\bibitem{GNRdis}
I. Romanovsky, C. Yannouleas and U. Landman, Phys. Rev. B
\textbf{83}, 045421 (2011); E. Prada, P. San-Jose and L. Brey,
Phys. Rev. Lett. \textbf{105}, 106802 (2010); C. Ritter, S. S.
Makler and A. Latg\'{e}, Phys. Rev. B \textbf{77}, 195443 (2008).


\bibitem{GNRee}
A. A. Shylau {\it et al.}, Phys. Rev. B. \textbf{82}, 121410(R)
(2010)

\bibitem{Sup}
See EPAPS Documents for supplementary material.

%


\bibitem{WilliamsAbanin}
J.R. Williams {\it et al.}, Phys. Rev. B \textbf{80}, 045408
(2009); D.A. Abanin and L. S. Levitov, Phys. Rev. B \textbf{78},
035416 (2008).




\end{thebibliography}
\end{document}